\documentclass[aps,twocolumn,pra,superscriptaddress,amsmath,showpacs,tightenlines]{revtex4-1}
\usepackage{bbm}
\usepackage{amssymb}
\usepackage{amsmath}
\usepackage{graphicx}
\usepackage{subfigure}
\usepackage{natbib}
\usepackage{epsfig}
\usepackage{amsfonts}
\usepackage{mathrsfs}
\usepackage{xcolor}
\usepackage[toc,page,title,titletoc,header]{appendix}

\begin{document}

\title{Engineering photonic dispersion relation and atomic dynamics in waveguide QED setup via long-range hoppings}

\author{Weijun Cheng}
\affiliation{College of Physics and Optoelectronics Engineering, Taiyuan University of Technology, Taiyuan 030024, China}
\affiliation{School of integrated circuits, Tsinghua University, Beijing 100084, China}
\author{Da-Wei Wang}
\affiliation{School of integrated circuits, Tsinghua University, Beijing 100084, China}
\author{Yang Xue}
\affiliation{School of integrated circuits, Tsinghua University, Beijing 100084, China}
\author{Zhihai Wang}
\email{wangzh761@nenu.edu.cn}
\affiliation{Center for Quantum Sciences and School of Physics, Northeast Normal University, Changchun 130024, China}
\author{Liantuan Xiao}
\email{xlt@sxu.edu.cn}
\affiliation{College of Physics and Optoelectronics Engineering, Taiyuan University of Technology, Taiyuan 030024, China}
\affiliation{Key Laboratory of Advanced Transducers and Intelligent Control System, Ministry of Education and Shanxi Province,
Taiyuan University of Technology, Taiyuan 030024, China}
\affiliation{Shanxi Key Laboratory of Precision Measurement Physics, Taiyuan University of Technology, Taiyuan 030024, China}
\affiliation{State Key Laboratory of Quantum Optics Technologies and Devices, Institute of Laser Spectroscopy, Shanxi University,
Taiyuan 030006, China}

\begin{abstract}
Non-trivial dispersion relations engineered in photonic waveguide for the precise control of atomic dynamics has recently attracted considerable attention.
Here, we study a system in which atoms are coupled to one-dimensional coupled-resonator waveguides with long-range hoppings.
By carefully engineering the $j$th-order nearest neighbor (${\rm JNN}$) hoppings between resonators, we construct linear dispersion relations with the chiral characteristic.
To quantify the degree of linearity, we analyze the propagation fidelities of Gaussian wave packets in these waveguides.
Furthermore, we demonstrate that such coupled-resonator waveguides can serve as versatile platforms for enabling directional atomic radiation and absorption.
Beyond linear dispersion relations, more general forms, including quadratic and cubic relations, can also be achieved through tailored ${\rm JNN}$-hoppings.
 Our study thus provides a unified framework for simulating atom-environment couplings with arbitrary dispersion relations.
\end{abstract}

\maketitle

\section{Introduction}

The construction of fully controllable quantum devices, such as quantum computers, quantum simulators, and quantum cryptographic systems, has sparked interest in exploring atomic dynamics~\cite{Gardiner}.
It has been well established that the radiation behavior of atoms is influenced not only by their intrinsic properties but also by their surrounding environment, a phenomenon known as the Purcell effect~\cite{Purcell}.
However, due to the weak coupling between atoms and the electromagnetic field environment, spontaneous emission dominates, making it experimentally challenging to observe the coherent dynamical behaviors of atoms. To overcome this limitation, considerable efforts have been devoted to enhancing light-matter interactions~\cite{Liu,WilsonRae,Marquardt,Rabl,Sarlette}.
A typical approach to enhance light-matter interactions is to confine the light field within a cavity, thereby reducing the mode volume to a sufficiently small scale~\cite{Rempe}. This method has led to significant advancements in the field of cavity quantum electrodynamics (QED)~\cite{Aoki}.

Another approach to control atomic dynamics involves coupling atoms to a one-dimensional (1D) waveguide~\cite{Sheremet}.
Compared with free space, 1D waveguides offer a structured continuum with well-defined dispersion relations, leading to strongly modified spontaneous emission processes~\cite{ParkJ,LodahlP}, collective decay~\cite{Tiranov}, and photon-mediated interactions~\cite{ChenJQ}.
The development of waveguide QED has been enabled across a broad range of platforms, including nanophotonic crystal waveguides~\cite{Goban,Liu2,Sipahigil}, surface plasmonic nanowires~\cite{Akimov}, optical nanofibers~\cite{KienFL,Rajasree,ZhangJ}, cold-atom waveguides~\cite{RosED,CorzoNV,ChangDE}, and superconducting transmission lines~\cite{ThomasH,Scigliuzzo,LiuY}. These systems have enabled the observation of non-Markovian dynamics~\cite{photon1,photon2}, bound states in the continuum~\cite{Stillinger,Marinica,Molina,Calajo, Qiu}, super- and subradiance~\cite{Kim,DincF}, and strong cooperative effects among multiple emitters~\cite{Wang}.

A major milestone in this field is the realization of chiral waveguides, where quantum emitters couple differently to left- and right-propagating modes.
 Over the past decade, a wide variety of mechanisms, such as spin-momentum locking in photonic crystal waveguides and nanofibers~\cite{PetersenJ,SuarezForero,Lodahl}, transverse spin-orbit interactions in plasmonic structures~\cite{GongSH}, and metasurface supported unidirectional modes~\cite{Gromyko}, have been developed to achieve chirality and enable unidirectional chiral emission.
Chiral light-matter interactions not only give rise to new topological photonic effects~\cite{Mayer} but also introduce an additional control dimension for quantum devices~\cite{Sukhov,RodríguezFortuno,Scheel,Kalhor,Lodahl}, and can enable deterministic quantum state transfer between distant qubits~\cite{Khandelwal,Stannigel,Cirac}.

While significant progress has been made in realizing chiral waveguides in the optical domain, implementing chirality in the microwave regime remains generally challenging.
For example, in superconducting circuit systems, conventional transmission line structures are typically reciprocal and lack built-in chiral mechanisms such as spin-momentum locking or natural symmetry breaking~\cite{JoshiC}.
Nevertheless, many engineered microwave photonic systems, such as microwave resonator arrays, offer highly tunable coupling strengths, phases, and interaction ranges through advances in device design, controllable couplers, and synthetic gauge fields~\cite{Gu-PR,Roushan,Hacohen-Gourgy,Saxberg,Carroll,Georgescu}.
In particular, microwave resonator arrays can naturally support controllable long-range interactions, enabling the design of customized dispersion relations and synthetic photonic lattices~\cite{DengX,Tennant,Kuster}. These advantages open new avenues for realizing programmable photonic environments.

The preceding analysis naturally leads to the question of what novel effects may arise when atoms are coupled to such programmable photonic environments.
In this work, we study the dynamics of atoms coupled to the coupled resonator waveguides with the $j$th-order nearest neighbor (${\rm JNN}$) hoppings.
We first show how to construct chiral linear dispersion relations in the waveguides by engineering the ${\rm JNN}$-hoppings.
The dispersion effects on Gaussian wave-packet propagation are then analyzed through the calculation of the propagating fidelities.
We also analytically and numerically calculate the directional radiation and absorption of atoms under various atom-waveguide coupling configurations.
Finally, we show that arbitrary dispersive behaviors, including quadratic and cubic dispersion relations, can be realized through appropriately engineered ${\rm JNN}$-hoppings between resonators.

\section{Results}
\label{symmetric}

\subsection{Model and Hamiltonian}

\begin{figure}[htbp]
\centering
 \includegraphics[width=8cm]{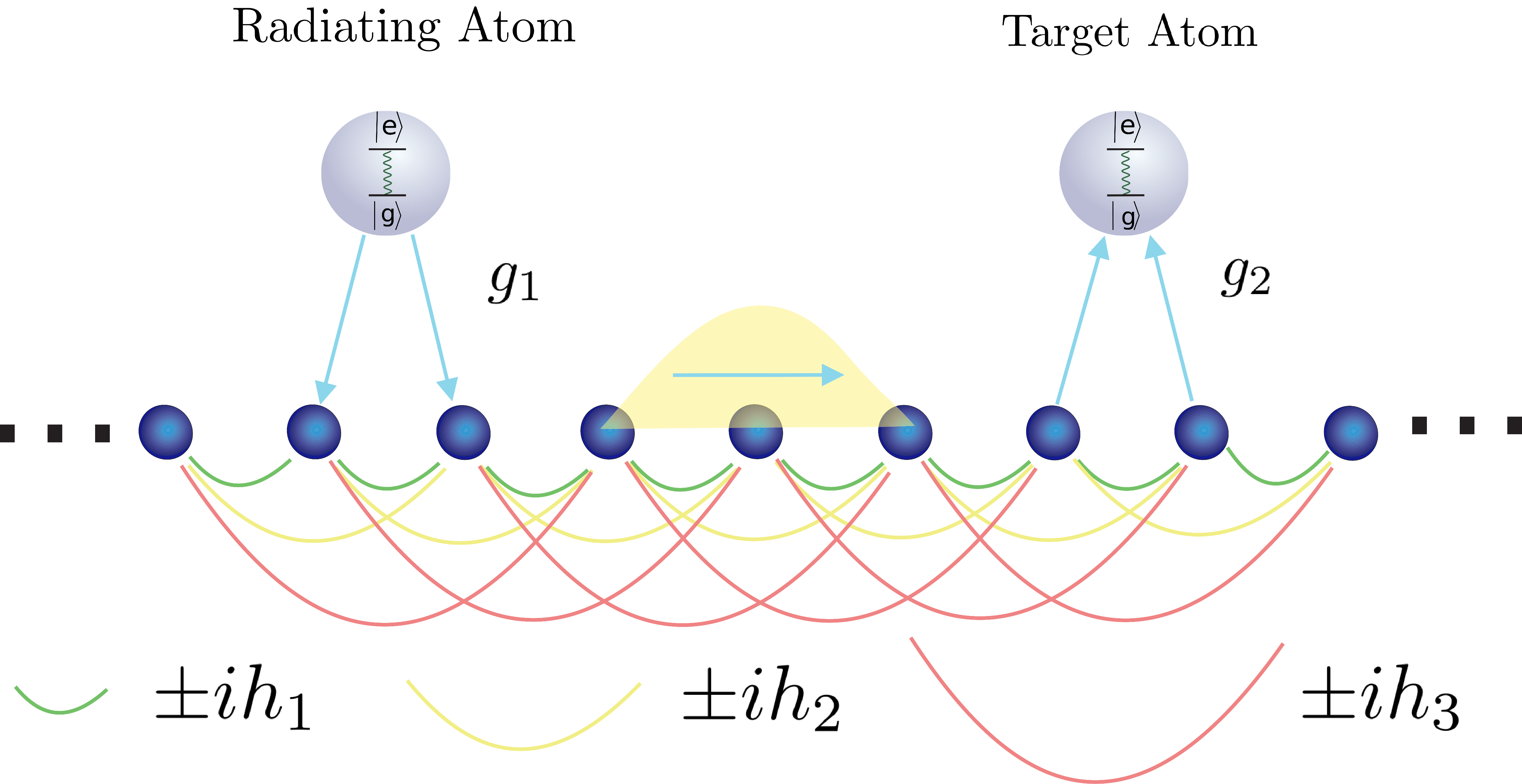}
\caption{Schematic diagram for the 1D coupled resonator waveguide with the long-range hoppings coupled to the two atoms.}
\label{model2}
\end{figure}

As schematically shown in Fig.~\ref{model2}, we consider a one-dimensional (1D) coupled resonator waveguide with the long-range hoppings coupled to the multiple atoms. Hereafter, we assume $\hbar=1$. The Hamiltonian of the whole system is
 \begin{eqnarray}
H=H_{{\rm JNN}}+H_I+\sum_n \omega_n |e\rangle_n\langle e|,
\label{H}
\end{eqnarray}
where $\omega_n$ is the transition frequency of the $n$-th atom between the ground state $|g\rangle$ and excited state $|e\rangle$.
The Hamiltonian $H_{{\rm JNN}}$ of the waveguide is given by
 \begin{eqnarray}
H_{{\rm JNN}}=\sum_{l=1}^L \left\{\omega_0 a_l^{\dag}a_l-\sum_{j=1}^{J}[h_je^{i\theta_j}a_l^{\dag}a_{l+j}+{\rm H.c}]\right\}.
\label{HJNN}
\end{eqnarray}
We assume that all resonators have the same frequency $\omega_0$.
$a_l$ is the annihilation operation of the $l$-th resonator and $h_j$ is the $j$th-order nearest neighbor(${\rm JNN}$) hopping strength.
$\theta_j$ denotes the phase of the ${\rm JNN}$-hoppings.
$L$ is the total number of resonators and $J$ is the maximum distance of the ${\rm JNN}$-hoppings.
 We apply the Fourier transform $a_l=\sum_k a_k e^{ikl}/\sqrt{L}$ to the Hamiltonian $H_{{\rm JNN}}$ of the waveguide under the periodical boundary condition.
 Then the Hamiltonian $H_{{\rm JNN}}$ is expressed as
 \begin{eqnarray}
H_{{\rm JNN}}(k)=\omega(k)a_k^{\dag}a_k,
\label{H0PK}
\end{eqnarray}
in the momentum space with
 \begin{eqnarray}
\omega(k)=\omega_0-\sum_{j=1}^J\left[2h_j\cos(jk+\theta_j)\right].
\label{omegak}
\end{eqnarray}
The interaction Hamiltonian $H_I$ is
 \begin{eqnarray}
H_I=\sum_{n=1}^N g_n(\sigma_n^{+}a_{l_n}+{\rm H.c}),
\label{HI}
\end{eqnarray}
where $\sigma_{n}^{+}=|e\rangle_n\langle g|$ ($\sigma_{n}^{-}=|g\rangle_n\langle e|$) is the raising (lowering) operator for the $n$-th atom located at lattice site $l_n$ and $g_n$ is the coupling strength.

\subsection{Chiral linear dispersion relations}

In the following, we demonstrate that chiral dispersion relations can be achieved by appropriately engineering the long-range hoppings in the coupled-resonator waveguide and utilize them to achieve directional radiation and complete absorption of atoms as shown in Fig.~\ref{model2}.
We emphasize that the term ``chiral" refers to a dispersion relation that lacks mirror symmetry with respect to $k=0$, that is $\omega(k)\neq\omega(-k)$.
To realize such a chiral linear dispersion, one must appropriately select the phase $\theta_j$ so that the resulting function becomes non-symmetric.
Here, by setting $\theta_j=\pi/2$,  the antisymmetric function with the sine-type formation can be generated
 \begin{eqnarray}
\omega(k)=\sum_k\left\{\omega_0+\sum_{j=1}^J\left[2h_j\sin(jk)\right]\right\}.
\label{omegak1}
\end{eqnarray}
It can be seen that the introduction of the phase $\theta_j$ modifies the functional form of the dispersion relation.
Next, we perform a Taylor expansion of the dispersion relation $\omega(k)$ around $k=0$ and retain it up to the $J$-level series, then Eq.\eqref{omegak1} can be express as
 \begin{eqnarray}
\omega(k)\approx\sum_k\left\{\omega_0+\sum_{i=1}^{J}C_i k^{2i-1}\right\},
\label{omegak1a}
\end{eqnarray}
with
 \begin{eqnarray}
C_i=\sum_{j=1}^J\frac{2h_j(-1)^{i+1}j^{2i-1}}{(2i-1)!}.
\label{Cn}
\end{eqnarray}

\begin{figure}[htbp]
\centering
  \includegraphics[width=8cm]{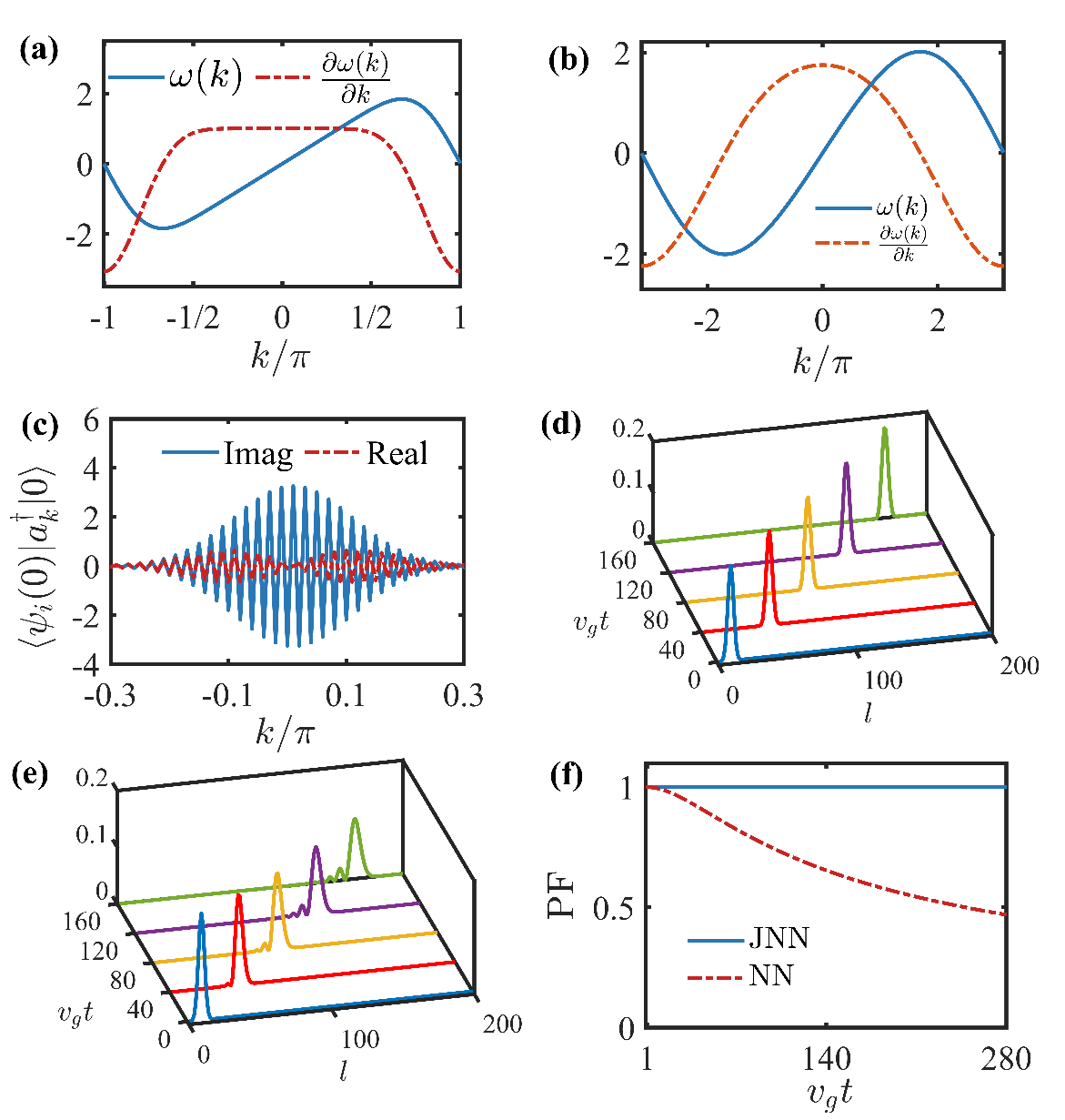}
\caption{(a) and (b) The linear dispersion relation $\omega(k)$ and its group velocity $v=\partial \omega(k)/\partial k$ versus $k$ for the waveguide with ${\rm JNN}$-hoppings and ${\rm NN}$-hoppings, respectively.
(c) The real part ${\rm Re}(\langle\psi_i(0)|a^{\dag}_{k}|0\rangle)$ and imaginary part ${\rm Im}(\langle\psi_i(0)|a^{\dag}_{k}|0\rangle)$ in the momentum space  for the  wave packet in Eq.~(\ref{wavepacket0}).
(d) and (e) The photon population distribution of $|\langle \psi_f(t)|a^{\dag}_{l}|0\rangle|^2$ for different time $t$ in the waveguide with ${\rm JNN}$-hoppings and ${\rm NN}$-hoppings, respectively. (f) The propagating fidelity ${\rm PF}$ of the initial wave packet $|\psi_i(0)\rangle$ versus time $t$. The parameters are set as: $L=300$, $\omega_0=0$, $\sigma=3$, $l_0=10$, $k_0=0$, $J=5$. $h_1=5v_g/6$, $h_2=-5v_g/21$, $h_3=5v_g/84$, $h_4=-5v_g/504$, $h_5=v_g/1260$ for (a), (d), (f). $h_1=v_g/2$, $h_2=0$, $h_3=0$, $h_4=0$, $h_5=0$ for (b), (e), (f).}
\label{gauss1}
\end{figure}

We expect that the dispersion relation exhibits a large linear region around $k\sim 0$.
From the expansion of $\omega(k)$, it can be seen that $\omega(k)$ generally contains terms of various orders of $k$.
To construct a linear dispersion relation, we need to retain the first-order terms and set the higher-order terms to zero.
Therefore, $C_i$ must satisfy the condition $C_1=v_g$ and $C_i=0$ for $2\leq i\leq J$, which is equivalent to a homogeneous linear system of $J$ variables.
When $J=5$, the solutions of the homogeneous linear system are $h_1=5v_g/6$, $h_2=-5v_g/21$, $h_3=5v_g/84$, $h_4=-5v_g/504$ and $h_5=v_g/1260$.
In this case, we plot the dispersion relation $\omega(k)$(the
solid blue line) and the group velocity $v=\partial \omega(k)/\partial k$ (the red dotted line) versus the wave vector $k$ as shown in Fig.~\ref{gauss1}(a), respectively.
Here, $\omega(k)$ is an odd function that exhibits a significant linear region within $k \in (-\pi/2, \pi/2)$. Correspondingly, the group velocity is depicted as a horizontal straight line in the region $k \in (-\pi/2, \pi/2)$.
 For comparison, we plot in Fig.~\ref{gauss1}(b) the dispersion relation and group velocity curve for the case with only nearest-neighbor (${\rm NN}$) hoppings. In this case, the hopping strengths are set to $h_1=v_g/2$, $h_2=0$, $h_3=0$, $h_4=0$ and $h_5=0$.
Here, the dispersion relation $\omega(k)=\omega_0+v_g\sin(k)$ exhibits a distinctly nonlinear character.

Waveguides with linear dispersion relations possess a strong capability for high-fidelity wave-packet transmission.
Thus, to characterize the degree of linearization of the waveguide’s dispersion relation, we study the dispersion effects of Gaussian wave packets localized around $k\sim 0$ in momentum space.
From the comparison of Fig. \ref{gauss1} (a) and (b), one sees that the inclusion of ${\rm JNN}$-hopping produces a much broader linear spectrum in the vicinity of
$k\sim 0$, thereby offering significantly improved wave packet transport.
The general form of the Gaussian wave packets~\cite{YWang} in the waveguide can be expressed as
 \begin{eqnarray}
|\psi_i(0)\rangle=\frac{1}{(2\sigma^2\pi)^{1/4}}\sum_{l=1}^L e^{-(l-l_0)^2/(2\sigma^2)}e^{ik_0l} a_{l}^{\dag}|0\rangle,
\label{wavepacket0}
\end{eqnarray}
where $l_0$ and $k_0$ is the center of the real space and momentum space of wave packet, respectively.
In Fig.~\ref{gauss1}(c), we numerically plot the distribution of the initial state $|\psi_i(0)\rangle=(2\sigma^2/\pi)^{1/4}\sum_k e^{-i(k-k_0)l_0}e^{-2\sigma^2(k-k_0)^2}a_{k}^{\dag}|0\rangle$ in momentum space.
We now study the evolution of the Gaussian wave packet $|\psi_i(0)\rangle$ expressed in Eq.~\eqref{wavepacket0} versus the time $t$.
 The final state $|\psi_f(t)\rangle=\exp(-iH_{\rm JNN}t)|\psi_i(0)\rangle$, as depicted in Fig.~\ref{gauss1} (d), remains localized in position space and propagates to the expected location with the group velocity $v_g$. This phenomenon arises from the compact distribution of wave packets in momentum space as shown in Fig.~\ref{gauss1}(c) and the modulated dispersion relation as shown in Fig.~\ref{gauss1} (a).
For the case with only ${\rm NN}$-hoppings, the Gaussian wave packet undergoes noticeable spreading during propagation, making it difficult to preserve its initial shape, as shown in Fig.~\ref{gauss1} (e).
 This phenomenon stems from the nonlinear character of the dispersion relation (Fig.~\ref{gauss1} (b)), which gives rise to mode-dependent group velocities and consequently causes the wave packet to spread as it propagates.

To quantify the distortion of the wave packet during propagation, we define the propagating fidelity (${\rm PF}$) as the fidelity between the evolved and ideal states
~\cite{ChengWJ}
 \begin{eqnarray}
{\rm PF}=|\langle \psi_f(t)|D(t)|\psi_i(0)\rangle|^2,
\label{PF}
\end{eqnarray}
where $|\psi_f(t)\rangle$ is the evolved wave packet and $D(t)$ is the translational operation expressed as
 \begin{eqnarray}
D(t)=\sum_{l=1}^{L}a^{\dag}_{l+v_gt}a_{l}.
\label{D1}
\end{eqnarray}
Here, we set $v_gt$ to be an integer for convenience. The wave packet undergoes perfect transmission when ${\rm PF}=1$, whereas ${\rm PF}=0$ corresponds to complete distortion.
In Fig.~\ref{gauss1}(f), we plot the ${\rm PF}$ of the Gaussian wave packet versus time $t$ for the case of ${\rm JNN}$-hoppings (the solid blue line) and ${\rm NN}$-hoppings (the red dotted line).
For the ${\rm NN}$-hoppings, the ${\rm PF}$ exhibits a rapid decay. In contrast, with ${\rm JNN}$-hoppings the ${\rm PF}$ remains close to 1. This behavior originates from the wide linear spectral window of the chiral dispersion, within which nearly all momentum components of the Gaussian wave packet reside, thereby minimizing dispersive effects.

\subsection{Directional radiation and absorption of atoms}

Up to now, we have built a chiral linear dispersion relation in the waveguide by engineering the ${\rm JNN}$-hoppings.
In this section, we now turn to study a specific application, that is, the directional radiation of the atom and the absorption of the target atom.

To achieve the directional radiation and absorption of the atoms, we consider the radiating atom and the target atom are coupled to two neighboring resonators ($l_n$ and $l_n+1$) of the waveguide as shown in Fig.~\ref{model2}.
Therefore, $H_I$ is replaced by $H_I^{\prime}$ as
 \begin{eqnarray}
H_I^{\prime}=\sum_{n=1,2} g_n(t)\left[\sigma_n^{+} (a_{l_n}+a_{l_n+1})+{\rm H.c}\right],
\label{HIp}
\end{eqnarray}
where the indices $l_1$ and $l_2$ represent the coupling position of the radiating atom and the target atom, respectively, and satisfy $l_1\ll l_2$.
$g_n(t)$ is the time-dependent variable.
Then the whole Hamiltonian of the system can be rewritten as
\begin{eqnarray}
H^{\prime}=H_{{\rm JNN}}+H_I^{\prime}+\sum_{n=1,2} \omega_{n} |e\rangle_n\langle e|.
\label{Hp}
\end{eqnarray}

To study the radiation dynamics of the atom and the absorption dynamics of the target atom, we assume that the single-excitation wave function at the time $t$ is given by
 \begin{eqnarray}
  \begin{split}
|\psi(t)\rangle=&e^{-i\omega_1t}b_1(t)\sigma_1^+|0\rangle
+e^{-i\omega_2t}b_2(t)\sigma_2^+|0\rangle \\
&+\sum_k e^{-i\omega(k)t}c_k(t)a_k^{\dag}|0\rangle,
\label{psit}
  \end{split}
\end{eqnarray}
where $b_1(t)$, $b_2(t)$ and $c_k(t)$ are the amplitudes of probability for the radiating atom, the target atom and the photon, respectively.
Performing some detailed calculations under the approximations ($\omega(k)\approx\omega_0+v_gk$ for $k\rightarrow 0$ and $\omega(k)\approx\omega_0- v_hk$ for $k\rightarrow \pm\pi$), the two-point coupling between the atoms and the waveguide will introduce photonic interference effects, which result in the atoms coupling only with the mode at $k\rightarrow0$ in the waveguide (see ``Methods'').  Here, $v_h$ is the group velocity of photons when they propagate to the left.
The approximate solutions of $b_1(t)$ and $b_2(t)$ under the condition $g_{1,2}\ll v_g$ can be expressed analytically as
 \begin{eqnarray}
b_1(t)\approx e^{-\int_0^td\tau\frac{2|g_1(\tau)|^2}{v_g}},
\label{b11}
\end{eqnarray}
 \begin{eqnarray}
  \begin{split}
b_2(t)\approx&-\int_{t_0}^tdt^{\prime}\frac{4g_2(t^{\prime})g_1(t^{\prime}-t_0)b_1(t^{\prime}-t_0)}{v_g}\\
&\times e^{-\int_0^{t^{\prime}-t_0}d\tau\frac{2|g_1(\tau)|^2b_1(\tau)}{v_g}}e^{-\int_{t^{\prime}}^t\frac{2|g_2(\tau)|^2}{v_g}}
\label{b22}
  \end{split}
\end{eqnarray}
under the boundary conditions $b_1(0)=1$ and $b_2(0)=0$.
Here, $t_0=(l_2-l_1)/v_g$.
Therefore, we emphasize that chirality is not only the asymmetry of the dispersion relation. Under the destructive interference effects, it can also allow the radiation photon from the atom to propagate unidirectionally in the waveguide (see ``Methods").

\subsubsection{Weak Coupling}

\begin{figure}[b]
\centering
  \includegraphics[width=9cm]{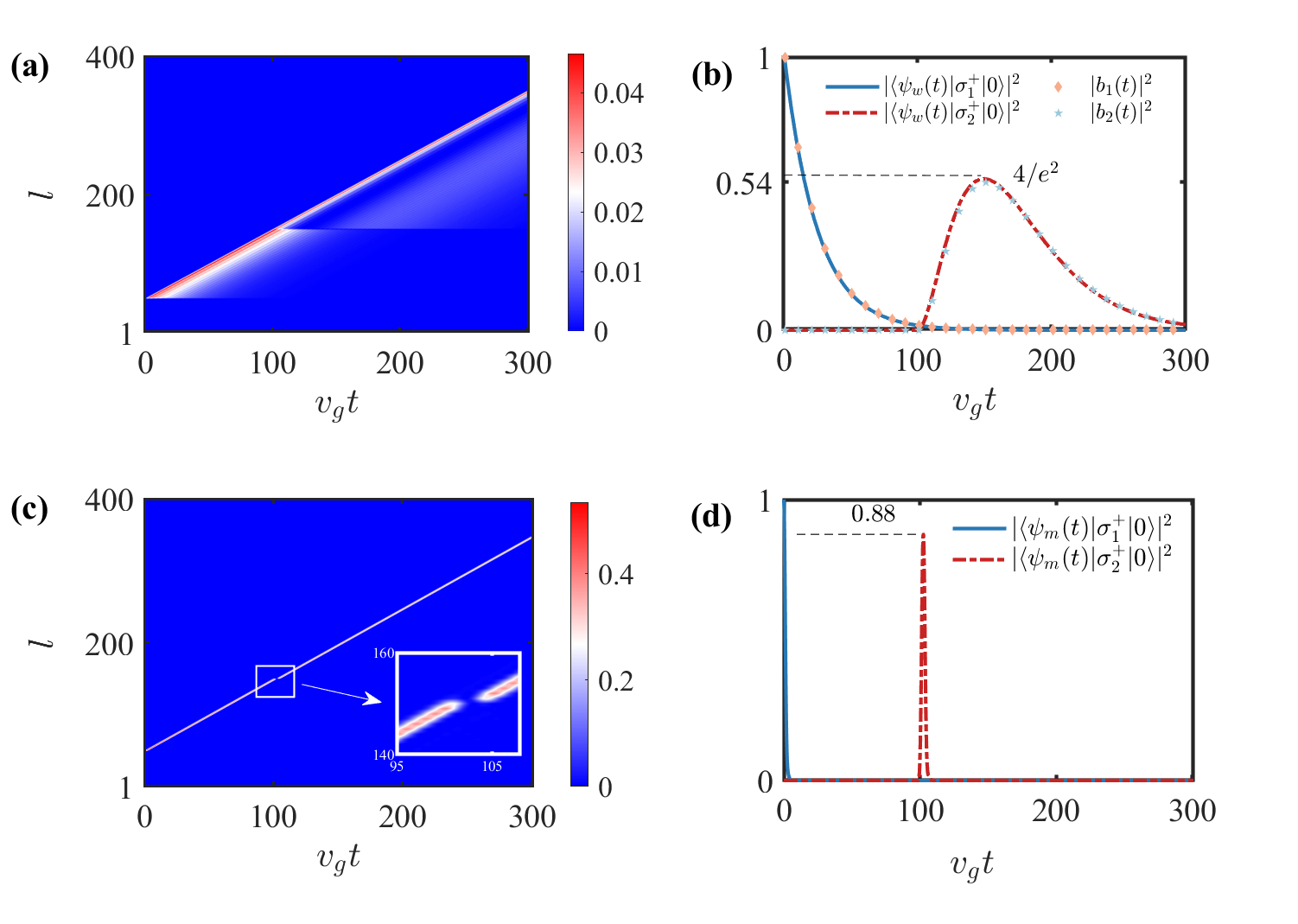}
\caption{(a) The photon population distribution of $|\langle \psi_{w}(t)|a^{\dag}_{l}|0\rangle|^2$ versus time $t$ in the waveguide. (b) The population distribution of $|\langle \psi_{w}(t)|\sigma_{1,2}^{+}|0\rangle|^2$ versus time $t$.
(c) The photon population distribution of $|\langle \psi_{m}(t)|a^{\dag}_{l}|0\rangle|^2$ versus time $t$ in the waveguide. (d) The population distribution of $|\langle \psi_{m}(t)|\sigma_{1,2}^{+}|0\rangle|^2$ versus time $t$. The parameters are set as: $L=400$, $l_1=50$, $l_1=150$, $g_1=g_2=0.5v_g$, $\omega_0=\omega_1=\omega_2$, $J=5$, $h_1=5v_g/6$, $h_2=-5v_g/21$, $h_3=5v_g/84$, $h_4=-5v_g/504$ and $h_5=v_g/1260$. $g_1=g_2=0.1v_g$ for (a) and (b). $g_1=g_2=0.5v_g$ for (c) and (d).}
\label{WC}
\end{figure}

\begin{figure}[htbp]
\centering
  \includegraphics[width=9cm]{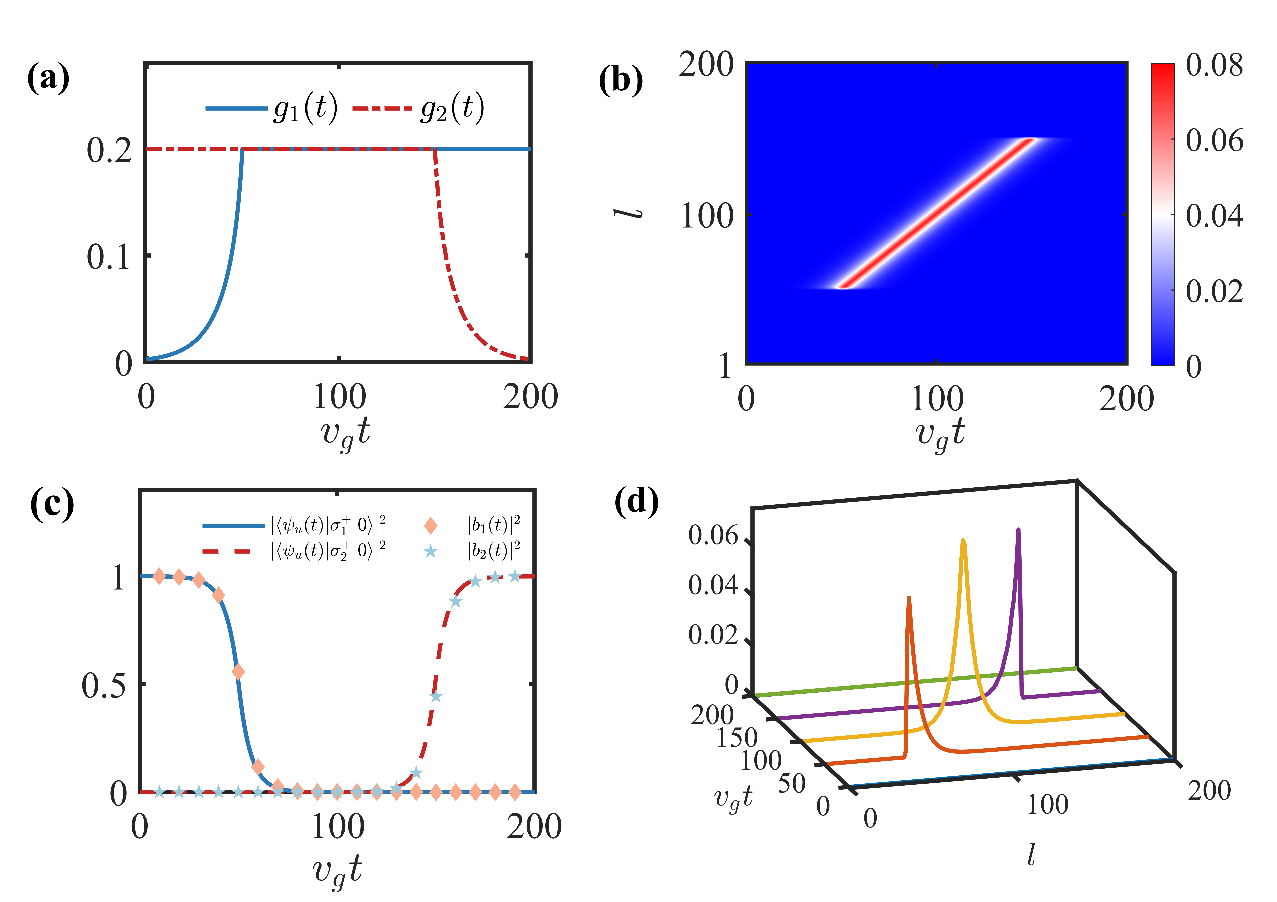}
\caption{(a) $g_1(t)$ and $g_2(t)$ versus time $t$. (b) The photon population distribution of $|\langle \psi_{u}(t)|a^{\dag}_{l}|0\rangle|^2$ versus time $t$ in the waveguide. (c) The population distribution of $|\langle \psi_{u}(t)|\sigma_{1,2}^{+}|0\rangle|^2$ versus time $t$.
(d)  The photon population distribution of $|\langle \psi_u(t)|a^{\dag}_{l}|0\rangle|^2$ for different time $t$ in the waveguide with ${\rm JNN}$-hoppings.
The parameters are set as: $L=200$, $l_1=50$, $l_1=150$, $\omega_0=\omega_1=\omega_2$, $g_{\rm max}=0.2$, $J=5$, $h_1=5v_g/6$, $h_2=-5v_g/21$, $h_3=5v_g/84$, $h_4=-5v_g/504$ and $h_5=v_g/1260$.}
\label{timeg}
\end{figure}

Up to this point, we have constructed the chiral linear dispersion relation, under which the atom couples exclusively to the modes in the vicinity of $k\rightarrow0$ in the waveguide.
For simplicity, we first study the radiation and absorption dynamics of the atom under weak coupling conditions.

When $g_{1,2}$ is time-independent, $b_1(t)$ and $b_2(t)$ in Eq.~\eqref{b11} and ~\eqref{b22} can be reduced to
 \begin{eqnarray}
b_1^{\prime}(t)\approx e^{-\frac{2|g_1|^2}{v_g}t}\theta(t),
\label{b11r}
\end{eqnarray}
 \begin{eqnarray}
  \begin{split}
b_2^{\prime}(t)\approx &-\frac{2g_2g_1^{\ast}}{|g_2|^2-|g_1|^2}\left\{e^{-\frac{2|g_1|^2}{v_g}(t-t_0)}\right.\\
&\left.-e^{-\frac{2|g_2|^2}{v_g}(t-t_0)}\right\}\theta(t-t_0),
\label{b22r}
  \end{split}
\end{eqnarray}
where $\theta(t-t_0)$ is the step function.
When $g_1=g_2=g$, $b_2(t)$ can be further simplified to
 \begin{eqnarray}
b_2^{\prime\prime}(t)\approx -\frac{4|g|^2}{v_g}(t-t_0)e^{-\frac{2|g|^2}{v_g}(t-t_0)}\theta(t-t_0).
\label{b22rr}
\end{eqnarray}

To illustrate the reabsorption dynamics of the target atom, we numerically calculate the evolution of system by $|\psi_w(t)\rangle=\exp(-iH^{\prime}t)\sigma_1^+|0,g\rangle$.
As shown in Fig.~\ref{WC}(a) and (b), we plot the photonic distribution $|\langle \psi_{w}(t)|a^{\dag}_{l}|0\rangle|^2$ of the waveguide and the population distribution $|\langle \psi_{w}(t)|\sigma_{1,2}^{+}|0\rangle|^2$ of the atoms, respectively.
The results indicate that the atomic radiation exhibits an exponential decay and the target atom can only absorb a portion of the emitted photon.
As shown in Fig.~\ref{WC}(b), we plot the analytical solutions of $|b_1^{\prime}(t)|^2$ and $|b_2^{\prime\prime}(t)|^2$ versus the time $t$ with quadrilaterals and pentagrams, respectively.
The agreement of numerical and analytical results shows the rationality of our approach.
Furthermore, we calculate the equation $\partial b_2^{\prime\prime}(t)\partial t=0$.
When $t=v_g/(2|g|^2)+t_0$, we can obtain the maximum $|b_{2}^{\prime\prime}(v_g/(2|g|^2)+t_0)|^2=4/e^2\approx0.54$ matching with the numerical result.

\subsubsection{Moderate coupling}

As the coupling strength increases to the moderate coupling ($g\sim v_g$), which is located in the region between weak coupling strength and strong coupling strength, the dynamics of the radiating atom will no longer exhibit exponential decay characteristics~\cite{Bernardis}.
In this case, we assume that the dynamic of system obeys $|\psi_m(t)\rangle=\exp(-iH^{\prime}t)\sigma_1^+|0,g\rangle$.
The population distribution $|\langle \psi_{m}(t)|\sigma_{1}^{+}|0\rangle|^2$ of the atom will experience rapid spontaneous emission as indicated by the solid blue line in Fig.~\ref{WC}(d).
The emitted photon $|\langle \psi_{m}(t)|a^{\dag}_{l}|0\rangle|^2$ will form a compact symmetrical wave packet that can propagate forward with a group velocity $v_g$ as shown in Fig.~\ref{WC}(c). In the moderate coupling regime, the wave packet is not only highly localized but also nearly symmetrical around its maximum, propagating through the lattice with minimal dispersion.

Next, we consider the absorption population distribution $|\langle \psi_{m}(t)|\sigma_{2}^{+}|0\rangle|^2$ of radiation photon by the target atom.
We found that when the radiation wave packet passes through the target atom, there is a brief disappearance of the photon in the waveguide as shown in inset of Fig.~\ref{WC}(c).
In Fig.~\ref{WC}(d), we can clearly observe that the target atom rapidly absorbs and re-emits the wave packet.
In comparison to weak coupling, the photon is reabsorbed with a probability approaching $P=0.88$ for a moderate coupled atom.
This high probability of absorption can be understood in terms of the symmetry of the emitted wave packet and its dispersionless propagation.

\subsubsection{Wave Packet Shaping}

While the probability approaching $P=0.88$ is already impressive, it would not be sufficient for a quantum computing application.
In fact, only waveforms that exhibit an exponentially increasing profile in the time domain can fully excite a two-level atom~\cite{Stannigel,Cirac,MichaelMurphy}.
To address this issue, researchers have proposed a wave packet shaping method. In this approach, by applying external Rabi control, the atom can emit a time-reversal symmetric wave packet, enabling complete absorption by the target atom.
Subsequently, the wave packet shaping was further extended to linear waveguide.
In this case, the coupling strengths $g_1(t)$ and $g_2(t)$ can be written as~\cite{Stannigel}
\begin{eqnarray}
g_1(t)=
\begin{cases}
g_{\rm max}
\sqrt{\frac{\exp(2g_{\rm max}^2(t-t_m)/v_g)}{2-\exp(2g_{\rm max}^2(t-t_m)/v_g)}} &  t<t_m\\
g_{\rm max} &  t\geq t_m
\end{cases},
\label{tm1}
\end{eqnarray}
\begin{eqnarray}
g_2(t)=
\begin{cases}
g_{\rm max} &  t\leq t_m+t_0\\
g_{\rm max}
\sqrt{\frac{\exp(2g_{\rm max}^2(t_m+t_0-t)/v_g)}{2-\exp(2g_{\rm max}^2(t_m+t_0-t)/v_g)}} &  t>t_m+t_0
\end{cases},\nonumber \\
\label{tm2}
\end{eqnarray}
where $g_{\rm max}$ is the maximum coupling strength between the atoms and waveguide.
$t_m$ ($t_m+t_0$) is the cutoff (starting) time for the exponential increase (reduce) of the time-dependent coupling strength $g_1(t)$ ($g_2(t)$).
In Fig.~\ref{timeg}(a), we plot the time-dependent coupling strengths $g_1(t)$ and $g_2(t)$ versus the time $t$.

Governed by the Hamiltonian in Eq.~\eqref{Hp} for $g_{1,2}(t)$ as shown in Fig.~\ref{timeg}(a), the dynamic evolution of system obeys $|\psi_{u}(t)\rangle=U(t)\sigma_1^+|0,g\rangle$, with the timing operator $U(t)=\mathcal {T}\exp\left[-i\int_0^td\tau H(\tau)\right]$.
As shown in Fig.~\ref{timeg}(b) and (c), we plot the photonic distribution $|\langle \psi_{u}(t)|a^{\dag}_{l}|0\rangle|^2$ of the waveguide and the population distribution $|\langle \psi_{u}(t)|\sigma_{1,2}^{+}|0\rangle|^2$ of the atoms, respectively.
The results show that the spatial distribution of the radiated photons in the waveguide exhibits a symmetric profile, as indicated by the yellow line ($v_gt=100$) in Fig.~\ref{timeg}(d).
This symmetry originates from the radiation of the atom located at $l_1=50$, whose emission process is symmetrically modulated by the time-dependent coupling strength $g_{1}(t)$ in Eq.~\eqref{tm1}, as indicated by the solid blue line in Fig.~\ref{timeg}(c).
Governed by $g_2(t)$ in  Eq.~\eqref{tm2}, the target atom can completely absorb the radiated photon, as indicated by the red dashed line in Fig.~\ref{timeg}(c).
In contrast to the radiation dynamics, the absorption dynamics of the target atom exhibit an antisymmetric character, manifested as an antisymmetric profile of $g_{2}(t)$ relative to $g_{1}(t)$ as shown in Fig.~\ref{timeg}(a).
As a result, when the radiated photon reach the position $l_2=150$ of the target atom, it suddenly vanish, as illustrated in Fig.~\ref{timeg}(b). This implies that the atomic excitation is completely transferred to the target atom after passing through the waveguide channel.
Moreover, as shown in Fig.~\ref{timeg}(c), we plot the analytical solutions of $|b_1(t)|^2$ and $|b_2(t)|^2$ (Eq.~\eqref{b11} and Eq.~\eqref{b22}) versus the time $t$ with quadrilaterals and pentagrams, respectively.
The agreement of numerical and analytical results shows the rationality of our approach.

\subsection{General forms of dispersion relations}
\label{General}

\begin{figure}[htbp]
\centering
 \includegraphics[width=9cm]{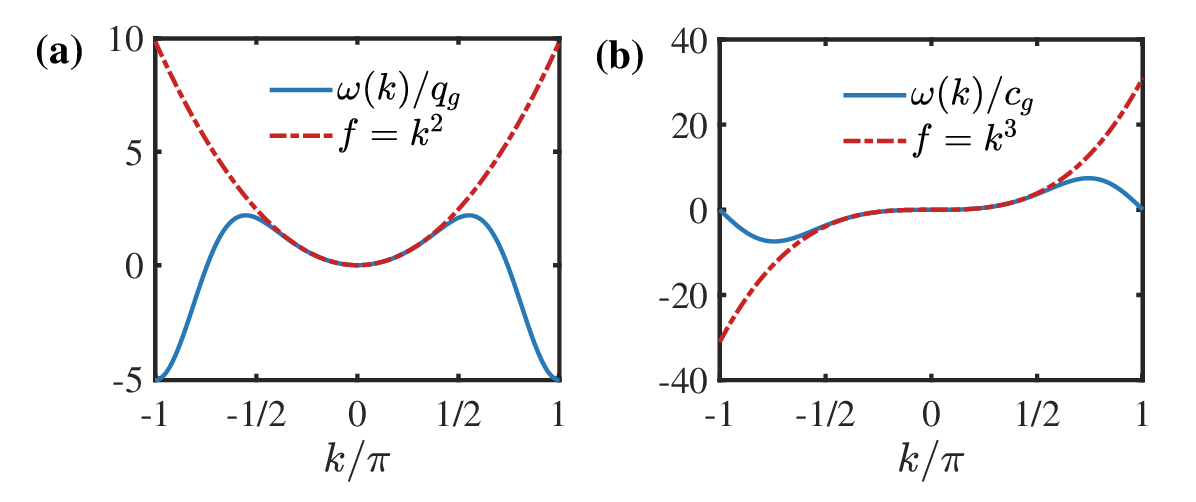}
\caption{ (a) The quadratic dispersion relation $\omega(k)$ and the function $f=k^2$ versus $k$.
 (b) The cubic dispersion relation $\omega(k)$ and the function $f=k^3$ versus $k$.
 The parameters are set as: $\omega_0=0$, $J=5$. $h_1=-826v_g/1069$, $h_2=586v_g/507$, $h_3=-242v_g/501$, $h_4=140v_g/1259$ and $h_5=-2v_g/177$ for (a).
 $h_1=843v_g/1091$, $h_2=-293v_g/507$, $h_3=-280v_g/1739$, $h_4=-35v_g/1259$ and $h_5=2v_g/885$ for (b).}
\label{qcenergy}
\end{figure}

Thus far, we have discussed how to generate a chiral dispersion relation and provided examples of its capabilities in manipulating atomic dynamics.
In fact, our scheme can equally well simulate waveguides with conventional symmetric dispersion relations (see ``Methods'').
By engineering the ${\rm JNN}$-hoppings, it is even possible to realize more general forms of waveguide dispersion relations.

For a more general case, we can set the coupling strength to $-\exp(i\theta_j)h_j$ in Eq.~\eqref{HJNN}. Under this case, the dispersion relation of the system can be expressed as
 \begin{eqnarray}
\omega(k)=\omega_0-\sum_{j=1}^J\left\{2h_j\left[\cos(\theta_j)\cos(jk)-\sin(\theta_j)\sin(jk)\right]\right\}.\nonumber \\
\label{omegakgen}
\end{eqnarray}
 Performing a Taylor expansion of the dispersion relation $\omega(k)$ around $k=0$, we can obtain
  \begin{eqnarray}
\omega(k)\approx\sum_k\left\{\omega_0-\sum_{i=0}^{J-1}D_i k^{2i}+\sum_{i=1}^{J}C_i k^{2i-1}\right\},
\label{omegakgenapp}
\end{eqnarray}
with
 \begin{eqnarray}
D_i=\sum_{j=1}^{J}\frac{2h_j\cos(\theta_j)(-1)^{i}j^{2i}}{(2i)!}
\label{DI}
\end{eqnarray}
and
 \begin{eqnarray}
C_i=\sum_{j=1}^J\frac{2h_j\sin(\theta_j)(-1)^{i+1}j^{2i-1}}{(2i-1)!}.
\label{CI}
\end{eqnarray}
This dispersion relation can include terms of all powers of $k$. Therefore, by adjusting the values of $D_i$ and $C_i$, any dispersion relation of the form  $\omega(k)=\sum_i \alpha_ik^i$($i=0,1...2J-1$) can be generated with the coefficient $\alpha_i$.

For example, a quadratic dispersion relation is a symmetric function that can be generated using Eq.~\eqref{omegakgenapp} by setting $\theta_j=0$.
Then, we eliminate all terms involving the wave vector $k$ except for the quadratic term. Thus, $D_i$ satisfy the conditions $D_1=-q_g$ and $D_i=0$ for $i=0$ and $2\leq i\leq J-1$, which is equivalent to a homogeneous linear system of $J$ variables.
Here, $q_g$ is the coefficient of the quadratic function.
When $J=5$, the solutions of the homogeneous linear system are $h_1=-826v_g/1069$, $h_2=586v_g/507$, $h_3=-242v_g/501$, $h_4=140v_g/1259$ and $h_5=-2v_g/177$.
In this case, we plot the dispersion relation $\omega(k)$(the solid blue line) and the function $f=k^2$(the red dotted line) versus the wave vector $k$ as shown in Fig.~\ref{qcenergy}(a).
In region $k\in(-\pi/2,\pi/2)$, the curve of $\omega(k)$ closely matches the function $f=k^2$. This indicates that $\omega(k)$ exhibits the characteristics of a quadratic function within the region $k\in(-\pi/2,\pi/2)$.
For another example, cubic dispersion relations are anti-symmetric functions that can be generated using Eq.~\eqref{omegakgenapp} with $\theta_j=\pi/2$. Similarly, we eliminate all terms except for the cubic term.
Thus, $C_i$ in Eq.~\eqref{CI} must satisfy the conditions $C_2=c_g$ and $C_i=0$ for $i=1$ and $3\leq i\leq J$ and a homogeneous linear system of $J$ variables is obtained. $c_g$ is the coefficient of the cubic function.
When $J=5$, the solutions of the homogeneous linear system are $h_1=843v_g/1091$, $h_2=-293v_g/507$, $h_3=-280v_g/1739$, $h_4=-35v_g/1259$ and $h_5=2v_g/885$.
In region $k\in(-\pi/2,\pi/2)$, the curve of $\omega^{\prime}(k)$ as shown the solid blue line in Fig.~\ref{qcenergy}(b) closely matches the function $f=k^3$ as shown the red dotted line in Fig.\ref{qcenergy}(b).

\section{Discussion}
\label{conclusion}

In summary, we have studied the interaction between atoms and a coupled resonator waveguide with the ${\rm JNN}$-hoppings.
By introducing an additional phases $\theta=\pi/2$, we construct chiral linear dispersion relations in the waveguide with the ${\rm JNN}$-hoppings.
We find that even after propagating through hundreds of lattice sites, the propagating fidelity of the Gaussian wave packet in this waveguide can still approach to $1$.
This exceptionally high propagating fidelity endows the waveguide with significant potential for the quantum state transfers.
We analytically and numerically calculate the directional radiation and absorption of the atoms under different coupling configurations.
By employing wave packet shaping, we successfully transfer the excitation of the radiating atom completely to the target atom.
Finally, we further develop the symmetric linear dispersion relations and clarify that our approach can be extended to generate any form of dispersion relation.

Our model may be implementable by using superconducting quantum circuits,
in which the superconducting qubits and $\rm LC$ resonator arrays act as the atoms and the waveguides, respectively~\cite{Gu-PR,Roushan,Hacohen-Gourgy,Saxberg}.
The existing superconducting circuits can achieve a coherence time of $100\mu s$, while the coupling strength between superconducting qubits and $\rm LC$ resonators can be modulated within $\rm 1MHz-1GHz$~\cite{Carroll,Georgescu}.
Therefore, the directional radiation and absorption of atoms mentioned in our work can theoretically be observed within the coherence time.
We hope that our work will contribute to the development of unconventional light-matter interactions.

\section{Methods}

\subsection{Dynamics of Atoms}\label{Appendix}

Governed by the Hamiltonian in Eq.~\eqref{Hp} and the assumption of the wave function in Eq.~\eqref{psit}, the dynamic equations can be obtained by $-i\partial|\psi(t)\rangle/\partial t=H^{\prime}|\psi(t)\rangle$, then we will have
 \begin{eqnarray}
\frac{\partial}{\partial t}b_1(t)&=&-i\frac{g_1(t)}{\sqrt{N}}\sum_ke^{i[\omega_1-\omega(k)]t}(e^{ikl_1}+e^{ik(l_1+1)})c_k(t),\nonumber \\
\label{eqsb1}
\end{eqnarray}
 \begin{eqnarray}
\frac{\partial}{\partial t}b_2(t)&=&-i\frac{g_2(t)}{\sqrt{N}}\sum_ke^{i[\omega_2-\omega(k)]t}(e^{ikl_2}+e^{ik(l_2+1)})c_k(t),\nonumber \\
\label{eqsb2}
\end{eqnarray}
 \begin{eqnarray}
\frac{\partial}{\partial t}c_k(t)&=&-i\frac{g_1^{\ast}(t)}{\sqrt{N}}e^{-i[\omega_1-\omega(k)]t}(e^{ikl_1}+e^{ik(l_1+1)})b_1(t)\nonumber \\
&&-i\frac{g_2^{\ast}(t)}{\sqrt{N}}e^{-i[\omega_2-\omega(k)]t}(e^{ikl_2}+e^{ik(l_2+1)})b_2(t).\nonumber \\
\label{eqsc}
\end{eqnarray}
By employing the explicit solution of Eq.~\eqref{eqsc}, we can obtain
 \begin{eqnarray}
c_k(t)&=&-i\int_0^td\tau\frac{g_1^{\ast}(\tau)}{\sqrt{N}}e^{-i[\omega_2-\omega(k)]\tau}(e^{ikl_1}+e^{ik(l_1+1)})b_1(\tau)\nonumber \\
&&-i\int_0^td\tau\frac{g_2^{\ast}(\tau)}{\sqrt{N}}e^{-i[\omega_2-\omega(k)]\tau}(e^{ikl_2}+e^{ik(l_2+1)})b_2(\tau),\nonumber \\
\label{eqsc1}
\end{eqnarray}
 and substituting Eq.~\eqref{eqsc1} into Eq.~\eqref{eqsb1} and ~\eqref{eqsb2}, we can derive
\begin{widetext}
 \begin{eqnarray}
\frac{\partial}{\partial t}b_1(t)&\approx&
-\frac{g_1(t)}{v_g}\sum_{x_1=l_1}^{l_1+1}\sum_{x_1=l_1}^{l_1+1}\int_0^te^{i[\omega_1-\omega(k)](t-\tau)}b_1(\tau)g_1^{\ast}(\tau)\delta\left[\tau-\left(t-\frac{x_1-x_1^{\prime}}{v_g}\right)\right]\nonumber \\
\frac{\partial}{\partial t}b_2(t)&\approx&
-\frac{g_2(t)}{v_g}\sum_{x_2=l_2}^{l_2+1}\sum_{x_2=l_2}^{l_2+1}\int_0^te^{i[\omega_2-\omega(k)](t-\tau)}b_2(\tau)g_2^{\ast}(\tau)\delta\left[\tau-\left(t-\frac{x_2-x_2^{\prime}}{v_g}\right)\right]\nonumber \\
&&-\frac{g_2(t)}{v_g}\sum_{x_2=l_2}^{l_2+1}\sum_{x_1=l_1}^{l_1+1}\int_0^te^{i[\omega_1-\omega(k)](t-\tau)}e^{i(\omega_2-\omega_1)t}b_1(\tau)g_1^{\ast}(\tau)\delta\left[\tau-\left(t-\frac{x_2-x_1}{v_g}\right)\right].
\label{eqsb1x}
\end{eqnarray}
\end{widetext}
where we assume $x_2\gg x_1$.  Here, we make approximations $\omega(k)\approx\omega^{\prime}+v_gk$ for $k\rightarrow 0$ and $\omega(k)\approx\omega- v_hk$ for $k\rightarrow \pm\pi$ with $v_h\gg v_g$ under the condition $g_{1,2}\ll v_g$ according to the dispersion relation in Fig.~\ref{gauss1}(a).
The Delta functions are originating from Fourier transform $\sum_k\exp[ik(x_n-x_m)-iv_gk(t-\tau)]/\sqrt{L}=\delta[x_n-x_m-v_g(t-\tau)]$ for $k\rightarrow0$, with $n=1,2$ and $m=1,2$.
However, when $k\rightarrow \pm\pi$ and $|x_n-x_n^{\prime}|=1$, $\sum_{x_n,x_n^{\prime}}\exp[ik(x_n-x_n^{\prime})-iv_hk(t-\tau)]=0$.
Therefore, the atoms are not coupled to the modes $k\rightarrow \pm\pi$ due to the destructive interference.
Finally, by solving Eqs.~\eqref{eqsb1x} and making the approximation $\delta\left[\tau-\left(t-(x_n-x_n^{\prime})/v_g\right)\right]\approx\delta\left[\tau-t\right]$, we can obtain $b_1$ and $b_2$ as shown in Eq.~\eqref{b11} and Eq.~\eqref{b22}.

\subsection{Construction of the symmetric linear dispersion relation}

In the above discussion, we have shown how engineered ${\rm JNN}$-hoppings enable the construction of chiral linear dispersion relations and, more generally, arbitrary dispersion profiles.
To further illustrate the universality of our scheme, we now construct a symmetric linear dispersion relation and demonstrate its capability for controlling atomic dynamics.
We first clarify that the term ``symmetric linear dispersion relation" refers to a dispersion relation that remains invariant under the inversion $k\rightarrow -k$, exhibiting mirror symmetry about $k=0$.
We set $\theta_j=0$ to generate a symmetric dispersion relation with a cosine-like form. Then we obtain
 \begin{eqnarray}
\omega(k)=\omega_0-\sum_{j=1}^J\left[2h_j\cos(jk)\right].
\label{omegak}
\end{eqnarray}
To construct a symmetric linear dispersion relation by engineering the ${\rm JNN}$-hoppings, we perform a Taylor expansion of the dispersion relation around $k\sim\pm\pi/2$.
Then Eq~.\eqref{omegak} can be expressed as
 \begin{eqnarray}
  \begin{split}
\omega(k)\mid_{k\sim\pm\pi/2}\approx & \omega_0-\sum_{i=0}^{[J/2]}\left\{A_i (k\pm\pi/2)^{2i}\right. \\
\mp B_i&(k\pm\pi/2)^{2i+1}\left.\right\},
\label{omegaka}
 \end{split}
\end{eqnarray}
with
 \begin{eqnarray}
A_i=\sum_{j=1}^J\frac{(-1)^i2h_j\cos(j\pi/2)j^{2i}}{(2i)!},
\label{An}
\end{eqnarray}
and
 \begin{eqnarray}
B_i=\sum_{j=1}^J\frac{(-1)^{i+1}2h_j\sin(j\pi/2)j^{2i+1}}{(2i+1)!}.
\label{An}
\end{eqnarray}
where $[J/2]$ represents the largest integer less than or equal to $J/2$. This order is the minimum required for linearization, as will be evident in the subsequent solution of the system of equations.

\begin{figure}[htbp]
\centering
  \includegraphics[width=9cm]{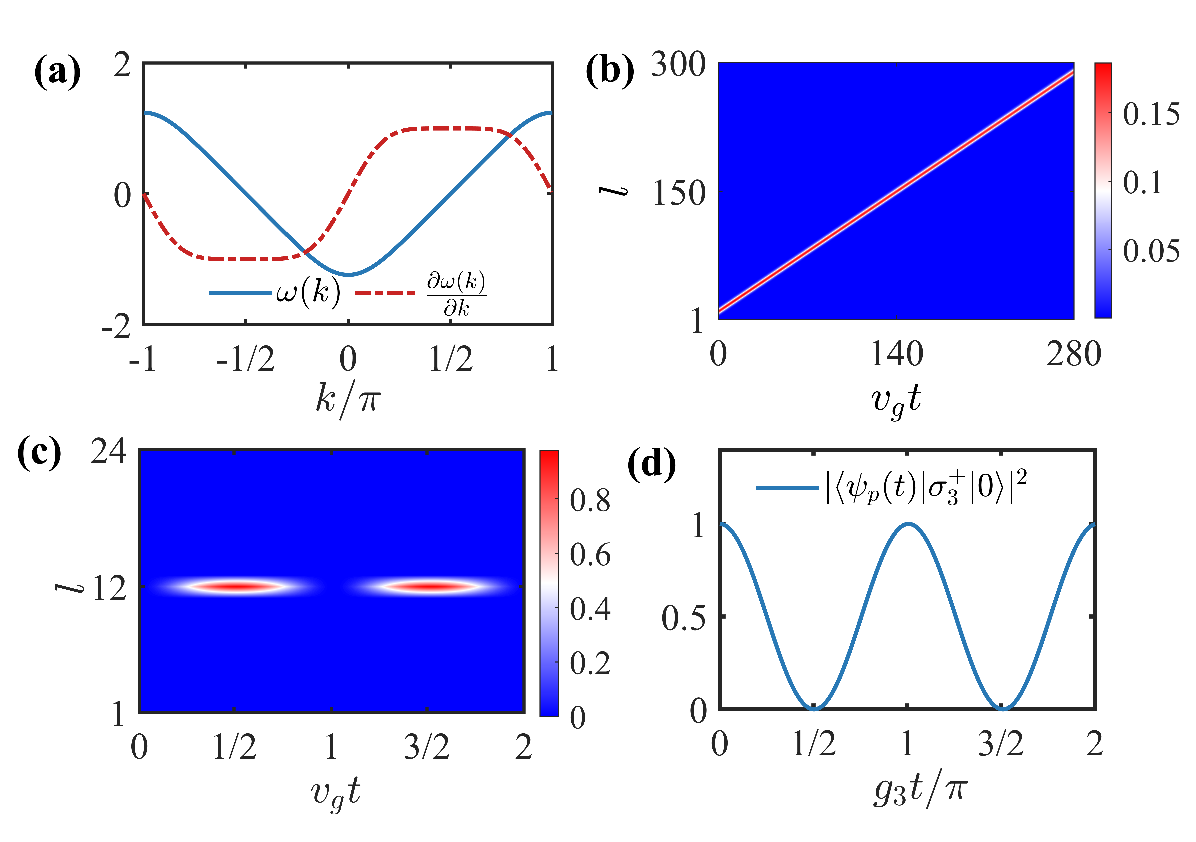}\nonumber\\
\caption{(a) The symmetric linear dispersion relation $\omega(k)$ and its group velocity $v=\partial \omega(k)/\partial k$ versus $k$.
(b) The photon population distribution of $|\langle \psi_f(t)|a^{\dag}_{l}|0\rangle|^2$ in the waveguide with the ${\rm JNN}$-hoppings.
 (c) The photon population distribution $|\langle \psi_p(t)|a^{\dag}_{l}|0\rangle|^2$ versus the time $t$ in the waveguide with ${\rm JNN}$-hoppings coupled to two atomic mirrors and a probe atom. (d) The population distribution $|\langle \psi_p(t)|\sigma_3^{+}|0\rangle|^2$ of the probe atom versus the time $t$.
The parameters are set as: $\omega_0=0$, $l_0=10$, $k_0=\pi/2$, $J=5$, $h_1=31v_g/53$, $h_2=0$, $h_3=25v_g/768$, $h_4=0$ and $h_5=3v_g/1280$. $\sigma=3$ and $L=1000$ for (b). $\omega_1=\omega_2=\omega_3=\omega_0$, $g_1=g_2=10v_g$, $g_3=0.1v_g$, $L=24$, $l_1=11$, $l_2=13$ and $l_3=12$ for (c) and (d).}
\label{Rabi}
\end{figure}

We expect that the dispersion relation exhibits a large linear region around $k\sim\pm\pi/2$.
From the expansion of $\omega(k)$, it can be seen that $\omega(k)$ generally contains terms of various orders of $k\pm\pi$.
To construct a linear dispersion relation, we need to retain the first-order terms and set the higher-order terms to zero.
 Therefore, $A_i$ and $B_i$ must satisfy the condition $A_i=0$ for $0\leq i\leq [J/2]$ and $B_0=v_g$, $B_i=0$ for $1\leq i\leq [J/2]$, which is equivalent to two homogeneous linear systems of $[J/2]+1$ variables.
$v_g$ is the group velocity in the linear regions.
For example, the solutions of the homogeneous linear system are $h_1=31v_g/53$, $h_2=0$, $h_3=25v_g/768$, $h_4=0$ and $h_5=3v_g/1280$ when $J=5$.
In this case, the dispersion relation of the waveguide is plotted with the solid blue line as shown in Fig.\ref{Rabi}(a). It is obvious that $\omega(k)$ is a even function of $k$ and satisfies $\omega(k)\sim\omega_0\pm v_g(k\mp\pi/2)$ around $k\sim\pm\pi/2$. It exhibits  a symmetric distribution around $k=0$ and a highly linear region around $k\sim\pm\pi/2$.
Correspondingly, the group velocity $v(k)=\partial \omega(k)/\partial k$ exhibits a horizontal line around $k\sim\pm\pi/2$ as shown the red dotted line in Fig.~\ref{Rabi}(a).
When the initial state is prepared as a Gaussian wave packet $|\psi_i(0)\rangle$ with its central mode at $k_0 = \pi/2$ as described by Eq.~\eqref{wavepacket0},
the dynamical evolution of the system is given by $|\psi_f(t)\rangle=\exp(-iH_{\rm JNN}t)|\psi_i(0)\rangle$. ($H_{\rm JNN}$ is the Hamiltonian of the waveguide with the symmetric dispersion relation.)
In Fig.~\ref{Rabi}(b), we plot the distribution of photons $|\langle\psi_f(t)|a_l^{\dag}|0\rangle|^2$ along the waveguide. Unsurprisingly, the photon distribution in the waveguide exhibits a high degree of localization and propagates to the specified location with the group velocity $v_g$.

\subsection{Cavity with Atomic Mirrors}

\begin{figure}[htbp]
\centering
  \includegraphics[width=8cm]{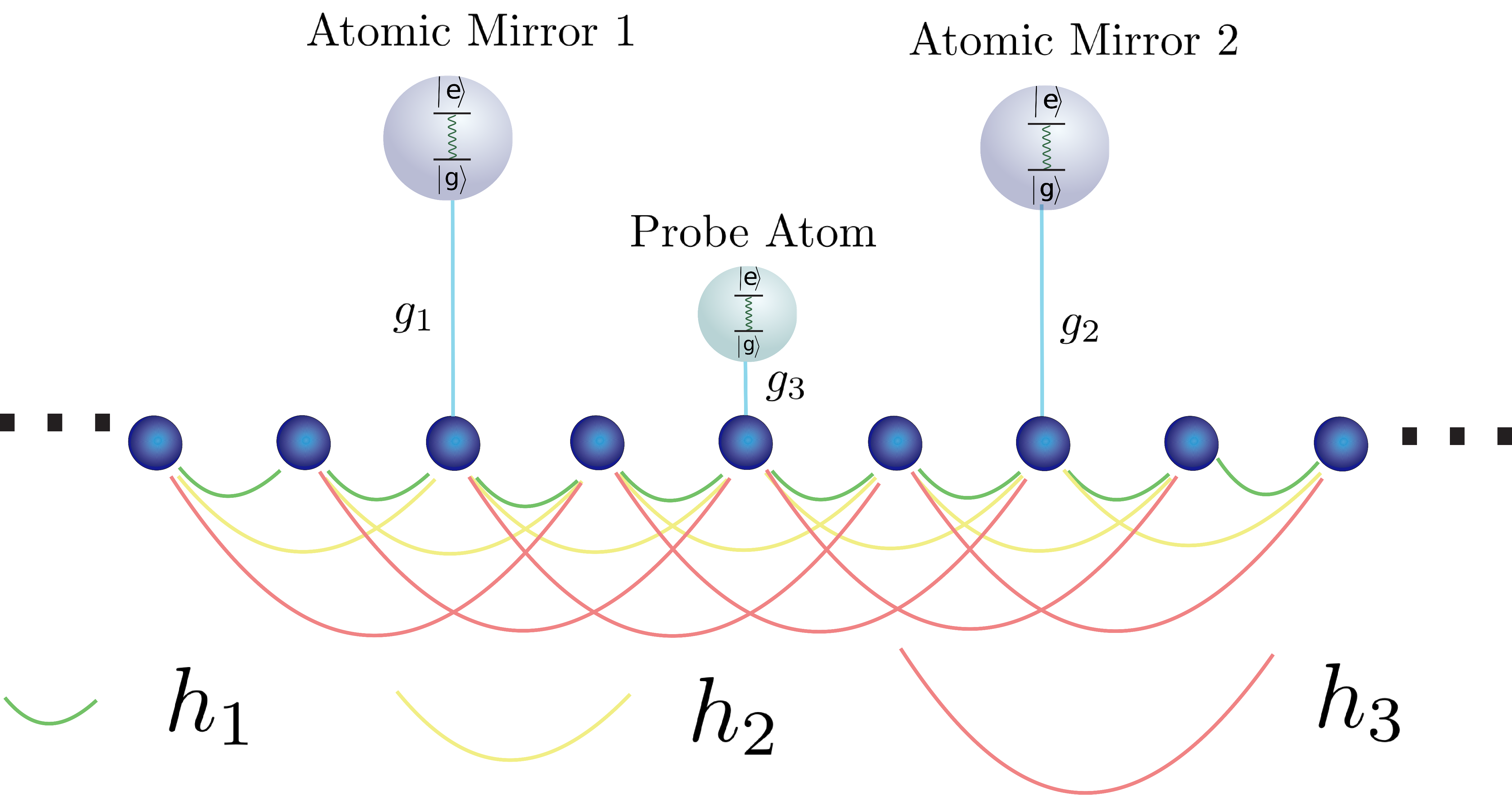}\nonumber\\
\caption{Schematic diagram for the multiple atoms coupling to the 1D coupled resonator waveguide with the long-range hoppings.}
\label{model}
\end{figure}

A series of quantum optical features have been discovered when one considers the replacement of a classical laser light by photons in cavity.
Fabry-P\'{e}rot cavity is one of the most typical quantum cavities, which consists of two parallel optical mirrors~\cite{Rempe,Aoki}.
Strong coupling between an atom and a waveguide allows the atom to serve as an efficient broadband reflector for the resonant modes. This has sparked considerable interest in optical microcavities that utilize atoms as microscopic mirrors~\cite{Chang,Nie}.
In this subsection, we will present an application: how to construct a cavity with atomic mirrors using the waveguide developed in the previous subsection.

We first will discuss the properties of scattering by the atom in the waveguide.
In general, the single-photon reflection spectrum of a atom in a waveguide with a symmetric linear dispersion relation exhibits Lorentzian line shapes~\cite{Nie} $R=|\Gamma/(\Delta+i\Gamma)|^2$, where $\Gamma=g_1^2/v_g=g_2^2/v_g$ is atomic decay rate and  $\Delta=\tilde{\omega}-\omega_1=\tilde{\omega}-\omega_2$ is the detuning between  the driving field and the atomic transition frequency.
Here, $g_n$ is the coupling strength between the waveguide and the $n$-th atom ($n=1,2$), while
$\tilde{\omega}$ and $\omega_{1(2)}$ represent the frequencies of the driving field and the
$1(2)$-th atom, respectively.
An atom functions as an ideal mirror when it completely reflects the incident field, which requires that the atom-waveguide coupling $g_{1(2)}$ satisfies $g_{1(2)}\gg\Delta_{1(2)}$.
Under this condition, a cavity can be formed by placing two strongly coupled atomic mirrors (atoms 1 and 2) in the waveguide, as schematically illustrated in Fig.~\ref{model}(a).

For a quantized cavity field, a characteristic feature is its ability to undergo Rabi oscillations with a two-level system.
Recent studies have shown that Rabi oscillations can be observed when the separation between the atomic mirrors is equal to the half-wavelength of the emitted photon $d=\lambda_0/2$ and the probe atom is placed at the center between the two atomic mirrors~\cite{Chang}.
In our model, the frequency of the probe atom is resonant with that of the resonators, i.e., $\omega_3=\omega_0$.
When the coupling strength satisfies $g_3\ll g_{1(2)}$, the frequency of photons emitted by the probe atom, $\tilde{\omega}$
, is distributed within $\tilde{\omega}\in(\omega_0-g_3,\omega_0+g_3)$.
This ensures that the detuning between the emitted photon and the atomic mirrors, $\Delta=\tilde{\omega}-\omega_1$, is much smaller than $g_{1(2)}$, enabling nearly perfect reflection of the photons by the atomic mirrors.
Accordingly, the photon modes $k$ emitted by the probe atom are distributed around $\pm\pi/2$, corresponding to a photon wavelength $\lambda_0=2\pi/k\approx4$.
Hence, the distance between the two atomic mirrors is set to $d=2$, and the probe atom is positioned precisely at their center.
Under these conditions, the time evolution of the probe atom is described by $|\psi_p(t)\rangle=\exp(-iHt)|\psi_p(0)\rangle$,
with the initial state $|\psi_p(0)\rangle=\sigma_3^{+}|0\rangle$ where $|0\rangle$ denotes the ground state of the entire system.
Figures~\ref{Rabi}(c) and \ref{Rabi}(d) show the evolution of the photon probability within the atomic cavity and the excitation probability of the probe atom, respectively.
Clear Rabi oscillations can be observed, characterized by $|\langle \psi_p(t)|\sigma_3^{+}|0\rangle|^2\sim \cos^2(\Omega t/2)$, with the Rabi frequency $\Omega=2g_3$.
For a single-mode optical cavity, if the intracavity photons leak to the outside (e.g., through the cavity mirror), the number of intracavity photons will decay exponentially $\langle n\rangle\sim e^{-\kappa t}$, with the  dissipation rate $\kappa$~\cite{Scully}.
For the cavity with atomic mirrors, the optical field will leak out through the atomic mirror. Roughly speaking, the average number of photons in the cavity is $\langle n\rangle\sim r^{v_g t/d}$, where $r$ is the reflection probability of the wave packet and $d$ is the separation between the atomic mirrors. $v_g t/d$ is the number of times the photon is reflected by the atomic mirror within time $t$.
By comparing the average photon number in the optical cavity and the cavity with atomic mirrors, we can obtain an effective dissipation rate for the cavity with atomic mirrors $\kappa=-v_g\ln (r/d)$.
These results demonstrate that the coupled resonator waveguide constructed in the previous subsection exhibits robust and stable performance, effectively functioning as a symmetric linear waveguide suitable for coherent light-matter interactions.

\begin{acknowledgments}
W.C. is supported by the funding from National Science Foundation of China (Grant No. 12405017) and China Postdoctoral Science Foundation (2024M761597).
D.W. is supported by the funding from China Postdoctoral Science Foundation (2025M773397).
Z.W. is supported by the funding from National Science Foundation of China (Grant No. 12375010).
L.X. is supported by the funding from Science and Technology Major Special Project of Shanxi Province (Grant No. 202201010101005); National Key Research and Development Program of China (Grant No. 2022YFA1404201).
\end{acknowledgments}


\begin{thebibliography}{10}


\bibitem{Gardiner} Gardiner, C. \& Zoller, P. \textit{The Quantum World of Ultra Cold Atoms and Light Book II: The Physics of Quantum Optical Devices}. (Imperial College Press, London, 2015).

\bibitem{Purcell} Purcell, E. M., Torrey, H. C., \&  Pound, R. V. Resonance absorption by nuclear magnetic moments in a solid. Phys. Rev. {\bf69}, 37 (1946).

\bibitem{Liu} Liu, H.-B.,  An, J.-H., Chen, C., Tong, Q.-J., Luo, H.-G. \& Oh, C. H. Anomalous decoherence in a dissipative two-level system. Phys. Rev. A {\bf87}, 052139 (2013).

\bibitem{WilsonRae}  Wilson-Rae, I., Nooshi, N., Zwerger, W., \&  Kippenberg, T. J., Theory of ground state cooling of a mechanical oscillator using dynamical backaction. Phys. Rev. Lett., {\bf99}, 093901 (2007).

\bibitem{Marquardt}  Marquardt, F.,  Chen, J. P., Clerk, A. A. \& Girvin, S. M. Quantum theory of cavity-assisted sideband cooling of mechanical motion. Phys. Rev. Lett. {\bf 99}, 093902 (2007).

\bibitem{Rabl}  Rabl, P., Shnirman, A., \& Zoller, P. Generation of squeezed states of nanomechanical resonators by reservoir engineering. Phys. Rev. B {\bf70}, 205304 (2004).

\bibitem{Sarlette} Sarlette, A., Raimond, J. M., Brune, M. \&  Rouchon, P. Stabilization of nonclassical states of the radiation field in a cavity by reservoir engineering. Phys. Rev. Lett. {\bf107} 010402 (2011).


\bibitem{Rempe} Rempe, G., Thompson, R. J., Kimble, H. J., \& Lalezari, R. Measurement of ultralow losses in an optical interferometer. Opt. Lett. {\bf17}, 5 (1992).

\bibitem{Aoki}  Aoki, T., Dayan, B., Wilcut, E., Bowen, W. P., Parkins, A. S., Kippenberg, T. J.,  Vahala, K. J., \& Kimble, H. J. Observation of strong coupling between one atom and a monolithic microresonator.  Nature {\bf443}, 7112 (2006).

\bibitem{Sheremet} Sheremet, A. S., Petrov, M. I., Iorsh, I. V., Poshakinskiy, A. V., \& Poddubny, A. N. Waveguide quantum electrodynamics: Collective radiance and photon-photon correlations. Rev. Mod. Phys. {\bf95}, 015002 (2023).

\bibitem{ParkJ} Park, J., Lee, K., Zhang, R.-Y., Park, H.-C. et al. Spontaneous Emission Decay and Excitation in Photonic Time Crystals. Phys. Rev. Lett. {\bf 35}, 133801 (2025).

 \bibitem{LodahlP} Lodahl, P., Mahmoodian, S. \& Stobbe S. Interfacing single photons and single quantum dots with photonic nanostructures. Rev. Mod. Phys. {\bf 87}, 347 (2015).

\bibitem{Tiranov} Tiranov, A., Angelopoulou, V. et al. ,Collective super- and subradiant dynamics between distant optical quantum emitters. Science {\bf 379}, 389 (2023).

 \bibitem{ChenJQ} Chen, J.-Q., Li, P.-B., G\'{o}mez-Le\'{o}n, A. \& Gonz\'{a}lez-Tudela, A. Photon-mediated interactions by Floquet photonic lattices. Phys. Rev. A {\bf 112}, 053710 (2025).

\bibitem{Goban} Goban, A., Hung, C.-L., Hood, J. D., Yu, S.-P., Muniz, J. A., Painter, O. \& Kimble, H. J. Superradiance for atoms trapped along a photonic crystal waveguide. Phys. Rev. Lett. {\bf115}, 063601 (2015).

\bibitem{Liu2}  Liu, Y. \& Houck, A. A. Quantum electrodynamics near a photonic bandgap, Nat. Phys. {\bf13}, 1 (2017).

\bibitem{Sipahigil}  Sipahigil, A., Evans, R. E., Sukachev, D. D., Burek, M. J., Borregaard, J.,  Bhaskar, M. K., Nguyen, C. T., Pacheco, J. L., Atikian, H. A., Meuwly, C. et al, An integrated diamond nanophotonics platform for quantum-optical networks, Science {\bf354}, 6314 (2016).

\bibitem{Akimov}  Akimov, A. V., Mukherjee,  A., Yu, C. L., Chang, D. E., Zibrov, A. S., Hemmer, P. R., Park, H. \&  Lukin, M. D. Generation of single optical plasmons in metallic nanowires coupled to quantum dots. Nature {\bf450}, 7168 (2007).

 \bibitem{KienFL} Kien, F. L., Hakuta, K., Reitz, D. Schneeweiss, P. \& Rauschenbeutel, A. Quantum dynamics of an atom orbiting around an optical nanofiber. Phys. Rev. A {\bf 87}, 063607 (2013).

\bibitem{Rajasree}  Rajasree, K. S., Ray, T., Karlsson, K., Everett, J. L. \&  Chormaic, S. N. Generation of cold Rydberg atoms at submicron distances from an optical nanofiber. Phys. Rev. Research {\bf 2}, 012038(R) (2020).

 \bibitem{ZhangJ} Zhang, J., Liu, K., Wang, P., Tong, L. \& Guo, X. Ultra-long-range optical pulling with an optical nanofibre. Nat. Commun. {\bf 16}, 7424 (2025).

 \bibitem{RosED} Ros, E. D., Cooper, N., Nute, J. \& Hackermueller L. Cold atoms in micromachined waveguides: A new platform for atom-photon interactions. Phys. Rev. Research {\bf 2}, 033098 (2020).


 \bibitem{CorzoNV} Corzo, N.V., Raskop, J., Chandra, A. et al. Waveguide-coupled single collective excitation of atomic arrays. Nature {\bf 566}, 359 (2019).

 \bibitem{ChangDE} Chang, D. E., Vuleti\'{c}, V. \& Lukin, M. D. Quantum nonlinear optics-photon by
photon. Nat. Photon. {\bf 8}, 685 (2014).

  \bibitem{ThomasH} Thomas, H., Marian, A., Chervyakov, A., Salmieri, S. D. \& Rubbia, C. Superconducting transmission lines-Sustainable electric energy transfer with higher public acceptance? Renew. Sust. Energ. Rev. {\bf 55}, 59 (2016).

  \bibitem{Scigliuzzo} Scigliuzzo, M., Calaj\`{o}, G., Ciccarello, F. et al. Controlling Atom-Photon Bound States in an Array of Josephson-Junction Resonators, Phys. Rev. X {\bf 12}, 031036 (2022).

  \bibitem{LiuY} Liu, Y., Houck, A. Quantum electrodynamics near a photonic bandgap. Nature Phys. {\bf 13}, 48 (2017).

\bibitem{photon1} Vega, I. de \&  Alonso, D. Emission spectra of atoms with non-Markovian interaction: Fluorescence in a photonic crystal. Phys. Rev. A {\bf 77}, 043836 (2008).

\bibitem{photon2} Hoeppe, U., Wolff, C., K\"{u}chenmeister, J., Niegemann, J., Drescher,  M., Benner H. \&  Busch, K. Direct Observation of Non-Markovian Radiation Dynamics in 3D Bulk Photonic Crystals. Phys. Rev. Lett. {\bf 108}, 043603 (2012).

\bibitem{Stillinger}  Stillinger, F. H. \&  Herrick, D. R. Bound states in the
continuum. Phys. Rev. A {\bf11}, 446 (1975).


\bibitem{Marinica}  Marinica, D. C., Borisov, A. G. \& Shabanov, S. V.
Bound States in the Continuum in Photonics. Phys. Rev.
Lett. {\bf100}, 183902 (2008).

\bibitem{Molina}  Molina, M. I., Miroshnichenko, A. E. \&  Kivshar, Y. S.
Surface Bound States in the Continuum. Phys. Rev. Lett.
{\bf108}, 070401 (2012).


\bibitem{Calajo}  Calajo, G., Fang, Y. L.,  Baranger, H. U. \& Ciccarello, F.
Exciting a Bound State in the Continuum through Multiphoton Scattering Plus Delayed Quantum Feedback. Phys. Rev. Lett. {\bf122}, 073601 (2019).

\bibitem{Qiu}  Qiu, Q.,  Wu, Y. \&  L\"{u}, X. Collective Radiance of Giant Atoms in Non-Markovian Regime. Sci. China Phys. Mech. Astron. {\bf66}, 224212 (2023).

\bibitem{Kim} Kim, Y., Lanuza, A. \& Schneble, D. Super- and subradiant dynamics of quantum emitters mediated by atomic matter waves. Nat. Phys. {\bf 21}, 70 (2025).

\bibitem{DincF} Dinc, F., Hayward, L. E. \& Branczyk A. M. Multidimensional super- and subradiance in waveguide quantum electrodynamics. Phys. Rev. Research {\bf 2},043149 (2020).


\bibitem{Wang}  Wang, Z.,  Jaako, T., Kirton, P. \&  Rabl, P. Supercorrelated Radiance in Nonlinear Photonic Waveguides. Phys. Rev. Lett. {\bf124}, 213601 (2020).

 \bibitem{PetersenJ} Petersen, J., Volz, J. \& Rauschenbeutel A. Chiral nanophotonic waveguide interface based on spin-orbit interaction of light. Science {\bf 346}, 67 (2014).

 \bibitem{SuarezForero} Su\`{a}rez-Forero, D. G., Mehrabad, M. J. \& Hafezi, C. V. M.  Chiral Quantum Optics: Recent Developments and Future Directions. Phys. Rev. X Quantum {\bf 6}, 020101 (2025).

 \bibitem{Lodahl} Lodahl, P., Mahmoodian, S., Stobbe, S. et al. Chiral quantum optics. Nature {\bf 541}, 473 (2017).

 \bibitem{GongSH} Gong, S.-H., Alpeggiani, F., Sciacca, B., Garnett, E. C. \& Kuipers, L. Nanoscale chiral valley-photon interface through optical spin-orbit coupling. Science {\bf 359}, 443 (2018).

\bibitem{Gromyko} Gromyko, D., An, S., Gorelik, S. et al. Unidirectional Chiral Emission via Twisted Bi-layer Metasurfaces. Nat. Commun. {\bf 15}, 9804 (2024).

\bibitem{Mayer} Mayer, N., Ayuso, D., Decleva, P. et al. Chiral topological light for detection of robust enantiosensitive observables. Nat. Photon. {\bf 18}, 1155 (2024).

\bibitem{Sukhov} Sukhov, S., Kajorndejnukul, V., Rezvani Naraghi, R. \& Dogariu, A. Dynamic
consequences of optical spin–orbit interaction. Nat. Photon. {\bf 9}, 809 (2015).

\bibitem{RodríguezFortuno} Rodríguez-Fortuno, F. J., Engheta, N., Martínez, A. \& Zayats, A. V. Lateral forces on circularly polarizable particles near a surface. Nat. Commun. {\bf 6}, 8799 (2015).

\bibitem{Scheel} Scheel, S., Buhmann, S. Y., Clausen, C. \& Schneeweiss, P. Directional
spontaneous emission and lateral Casimir–Polder force on an atom close to a
nanofiber. Phys. Rev. A {\bf 92}, 043819 (2015).

\bibitem{Kalhor} Kalhor, F., Thundat, T. \& Jacob, Z. Universal spin–momentum locked optical
forces. Appl. Phys. Lett. {\bf 108}, 061102 (2016).


 \bibitem{Khandelwal} Khandelwal, S., Chen,  W., Murch, K. W. \& Haack, G. Chiral Bell-State Transfer via Dissipative Liouvillian Dynamics. Phys. Rev. Lett. {\bf 133}, 070403 (2024).

\bibitem{Bernardis} D. D. Bernardis, F. S. Piccioli, P. Rabl, and I. Carusotto, Chiral Quantum Optics in the Bulk of Photonic Quantum Hall Syste, Phys. Rev. X Quantum {\bf 4}, 030306 (2023).

\bibitem{Stannigel} Stannigel, K., Rabl, P., Srensen, A. S., Lukin, M. D. \& Zoller, P. Optomechanical transducers for quantum-information processing. Phys. Rev. A {\bf84}, 042341 (2011).

\bibitem{Cirac}  Cirac, J. I., Zoller, P., Kimble, H. J. \& Mabuchi, H. Quantum State Transfer and Entanglement Distribution among Distant Nodes in a Quantum Network. Phys. Rev. Lett. {\bf78}, 3221 (1997).


\bibitem{JoshiC} Joshi, C., Yang, F. \& Mirhosseini, M. Resonance Fluorescence of a Chiral Artificial Atom. Phys. Rev. X {\bf 13}, 021039 (2023).


\bibitem{Gu-PR}X. Gu, A. F. Kockum, A. Miranowicz, Y. X. Liu, and F. Nori, Microwave photonics with superconducting quantum circuits, Phys. Rep.~\textbf{718-719}, 1 (2017).

\bibitem{Roushan} P. Roushan, C. Neill, J. Tangpanitanon, V. M. Bastidas, A. Megrant, R. Barends, Y. Chen, Z. Chen, B. Chiaro, A. Dunsworth, A. Fowler,B. Foxen, M. Giustina, E. Jeffrey, J. Kelly, E. Lucero, J. Mutus, M. Neeley, C. Quintana, D. Sank, A. Vainsencher, J. Wenner, T. White, H. Neven, D. G. Angelakis, J. Martinis, Spectroscopic signatures of localization with interacting photons in superconducting qubits, Science {\bf358}, 1175 (2017).

\bibitem{Hacohen-Gourgy} S. Hacohen-Gourgy, V. V. Ramasesh, C. D. Grandi, I. Siddiqi, and S. M. Girvin, Cooling and Autonomous Feedback in a Bose-Hubbard Chain with Attractive Interactions, Phys. Rev. Lett. {\bf115}, 240501 (2015).

\bibitem{Saxberg} R. Ma, B. Saxberg, C. Owens, N. Leung, Y. Lu, J. Simon, and D. I. Schuster, A dissipatively stabilized Mott insulator of photons, Nature {\bf566}, 51 (2019).

\bibitem{Carroll} M. Carroll, S. Rosenblatt, P. Jurcevic,  I. Lauer and, A. Kandala, Dynamics of superconducting qubit relaxation times, npj Quantum Inf. {\bf8}, 132 (2022).

\bibitem{Georgescu} I. M. Georgescu, S. Ashhab, and Franco Nori, Quantum simulation, Rev. Mod. Phys. {\bf86}, 153 (2014).


\bibitem{DengX} Deng, X., Zheng, W., Liao, X., Zhou, H., Ge, Y., Zhao, J., Lan, D., Tan, X., Zhang, Y. et al. Long-Range Interaction via Resonator-Induced Phase in Superconducting Qubits. Phys. Rev. Lett. {\bf 134}, 020801 (2025).

\bibitem{Tennant} Tennant, D. M., Dai, X., Martinez, A.J. et al. Demonstration of long-range correlations via susceptibility measurements in a one-dimensional superconducting Josephson spin chain. npj Quantum Inf. {\bf 8}, 85 (2022)

\bibitem{Kuster} K\"{u}ster, F., Brinker, S., Lounis, S. et al. Long range and highly tunable interaction between local spins coupled to a superconducting condensate. Nat. Commun. {\bf 12}, 6722 (2021).

\bibitem{YWang} Wang, Y., Zhang, Y., Zhang, Q., Zou,  B. \& Schwingenschlogl, U., Dynamics of single photon transport in a one-dimensional waveguide two-point coupled with a Jaynes-Cummings system. Sci. Rep. {\bf 6}, 33867 (2016).

\bibitem{ChengWJ} Cheng, W., Wang, Z. \& Liu, Y.-X. Controllable single-photon wave packet scattering in two-dimensional resonator array by a giant atom. 	arXiv:2410.20123.

\bibitem{MichaelMurphy} Murphy, M. \&  Montangms, S. Chiral Quantum Optics in the Bulk of Photonic Quantum Hall Systems. Phys. Rev. X Quantum {\bf4}, 030306 (2023).



\bibitem{Scully} Scully, M. O. \& Zubairy, M. S. \textit{Quantum Optics}. (Cam-bridge University Press, Cambridge, 1997).

\bibitem{Chang} D. E. Chang, L. Jiang, A. Gorshkov, and H. Kimble, Cavity QED with atomic mirrors, New J. Phys. {\bf14}, 063003 (2012).

\bibitem{Nie} W. Nie, T. Shi, Y. Liu, and F. Nori, Non-Hermitian Waveguide Cavity QED with Tunable Atomic Mirrors, Phys. Rev. Lett. {\bf131}, 103602 (2023).
%%%%%%%%%%%%%%%%%%



\end{thebibliography}
\end{document}